\documentclass[]{aa}

\usepackage{epsfig,graphicx,natbib}

\usepackage{longtable}
\usepackage[varg]{txfonts}

\newcommand{\felix}[1]{\textsc{F\small{ELIX}}}
\makeatother

\begin{document}


\title{Amplitude and frequency variations of oscillation modes\\ in the
pulsating DB white dwarf star KIC~08626021}

\subtitle{The likely signature of nonlinear resonant mode coupling}

\author{W.~Zong \inst{1,2}
\and S.~Charpinet \inst{1,2}
\and G.~Vauclair \inst{1,2}
\and N.~Giammichele \inst{3}
\and V.~Van~Grootel \inst{4}
}

\institute{Universit\'e de Toulouse, UPS-OMP, IRAP, Toulouse F-31400, France
\and CNRS, IRAP, 14 avenue Edouard Belin, F-31400 Toulouse, France\\ 
\email{[weikai.zong,stephane.charpinet,gerard.vauclair]@irap.omp.eu}
\and D\'epartement de Physique, Universit\'e de Montr\'eal, CP 6128, Succursale Centre-Ville,
Montr\'eal, QC H3C 3J7, Canada\\
\email{noemi@astro.umontreal.ca}
\and Institut d'Astrophysique et de G\'eophysique, Quartier Agora, 
All\'ee du 6 Ao\^ut 19c, 4000 Li\`ege, Belgium.\\
\email{valerie.vangrootel@ulg.ac.be}
}

\date {Received  / Accepted}

\titlerunning{Nonlinear resonant mode coupling in the pulsating DB white dwarf star KIC~08626021}
\authorrunning{Zong et al.}

\abstract
{The signatures of nonlinear effects affecting stellar oscillations are 
difficult to observe from ground observatories due 
to the lack of continuous high precision photometric data spanning extended 
enough time baselines. The unprecedented photometric quality and coverage 
provided by the {\sl Kepler} spacecraft offers new opportunities to 
search for these phenomena.}
{We use the {\sl Kepler} data accumulated on the pulsating DB white dwarf 
KIC\,08626021 to explore in detail the stability of its oscillation modes, 
searching in particular for evidences of nonlinear behaviors.}
{We analyse nearly two years of uninterrupted short cadence data, 
{concentrating on identified triplets caused by stellar 
rotation} 
that show intriguing behaviors during the course of the observations.}
{We find clear signatures of nonlinear {effects that could be} 
attributed to resonant mode 
coupling mechanisms. These couplings occur between the components of 
the triplets and can induce different types of behaviors. We first 
notice that a structure at 3681\,$\mu$Hz identified as 
a triplet in previous published studies is in fact forming a doublet with 
the third component being an independent mode. We find that a triplet 
at 4310\,$\mu$Hz and this doublet at 3681 $\mu$Hz (most likely the two 
visible components of an incomplete triplet) have clear periodic 
frequency and amplitude modulations typical of the so-called intermediate 
regime of the resonance, with time scales consistent 
with theoretical expectations. Another triplet at 5073\,$\mu$Hz is likely 
in a narrow transitory regime in which the amplitudes are 
modulated while the frequencies are locked. Using 
nonadiabatic pulsation calculations based on a model representative of 
KIC\,08626021 to evaluate the linear growth rates of the modes in the triplets, 
we also provide quantitative information that could be useful for future 
comparisons with numerical solutions of the amplitude equations.}
{{The observed modulations are the clearest hints of
nonlinear resonant couplings occurring in white dwarf stars identified so 
far}. These should resonate as a warning to projects aiming at measuring 
the evolutionary cooling rate of KIC\,08626021, and of white dwarf stars 
in general. Nonlinear modulations of the frequencies can potentially 
jeopardize any attempt to measure reliably such rates, unless they
can be corrected beforehand. These results should motivate further 
theoretical work to develop the nonlinear stellar pulsation theory.}


\maketitle
               
\section{Introduction}

The temporal variations of the amplitude and frequency of oscillation modes 
often seen, or suspected, in pulsating stars cannot be explained by the 
linear nonradial stellar oscillation theory \citep{un89} and must be 
interpreted in the framework of a nonlinear theory. It is believed that nonlinear 
mechanisms such as resonant mode couplings could generate such modulations, 
as, e.g., in the helium dominated atmosphere (DB) white dwarf star GD~358 
\citep{go98}. Resonant couplings are for instance predicted to occur when slow 
stellar rotation produces triplet structures whose component 
frequencies satisfy the relation $\nu_+ + \nu_- \sim 2\nu_0$, where $\nu_0$ 
is the frequency of the central $m=0$ mode.
The theoretical exploration of these mechanisms was extensively developed 
in \citet{bu95,bu97}, but was almost interrupted more than a decade ago 
because of the lack of clear observational evidence of such phenomena, 
due to the difficulty of capturing amplitude or frequency variations that 
occur on months to years timescales from ground based observatories. 
Nevertheless, the presence of resonant couplings within rotationally split 
mode triplets was proposed for the first time as the explanation for the 
frequency and amplitude long term variations observed in the {\sl GW Vir} 
pulsator PG~0122+200 \citep{va11} from successive campaigns on this object.
This suggests that pulsating white dwarfs could be among the best 
candidates to detect and test the nonlinear resonant coupling theory.

White dwarfs constitute the ultimate evolutionary fate expected for 
$\sim 98\%$ of the stars in our Galaxy. While cooling down, they cross 
several instability strips in which they develop observable nonradial $g$-mode 
oscillations. Among these, the helium atmosphere DB white dwarfs 
representing $\sim 20\%$ of all white dwarfs, are found to pulsate in the 
effective temperature range of 21,000\,K to 28,000\,K \citep{be99,fo08,wi08}.
All classes of pulsating white dwarfs are particularly valuable for probing 
their interior with asteroseismology, but it has also been proposed that 
hot DB pulsators with apparently stable modes could be used to measure their 
cooling rate, which is dominated by neutrino emission \citep{wi04}. The 
secular rates of change for the pulsation periods in hot DB pulsators is 
expected to be $\sim 10^{-13}$\,ss$^{-1}$, corresponding to a time scale of 
3$\times10^5$ years. However, this possibility could be seriously 
impaired by other phenomena affecting the pulsation frequencies on 
shorter timescales. Such variations in amplitude and 
frequency have indeed been suspected in several white dwarf stars 
(e.g.~PG~0122+200, \citealp{va11}; WD~0111+0018, \citealp{he13}; 
HS~0507+0434B, \citealp{fu13}), as stellar evolution theory cannot explain 
the variations with estimated timescales at least two orders of magnitude 
shorter than the expected cooling rates.
Nonlinear effects on stellar pulsations, including resonant mode coupling 
mechanisms could induce such modulations and need to be considered carefully
\citep{va13}.

In this context, observations from space of a multitude of pulsating stars 
including white dwarfs has open up new horizons. The {\sl Kepler} spacecraft 
monitored a 105 deg$^2$ field in the Cygnus-Lyrae region for nearly four 
years without interruption, obtaining unprecedented high quality photometric 
data for asteroseismology \citep{gi10}. 
These uninterrupted data are particularly suited to search for long term 
temporal frequency and amplitude modulations of the oscillation modes.

Among the 6 pulsating white dwarfs discovered in the {\sl Kepler} field, 
KIC~08626021 (a.k.a., WD J1929+4447 or GALEX~J192904.6+444708) is the 
only identified DB pulsator \citep{os11}. Based on the first month of short 
cadence (SC) {\sl Kepler} data, \citet{os11} estimated that this star has
an average rotation period $P_{\rm rot}\sim1.7$ days, derived from
the observed frequency spacings of 3 groups of $g$-modes interpreted as 
triplets due to rotation. Subsequent independent 
efforts to isolate a seismic model for KIC~08626021 from \citet{bi11} and 
\citet{co12} both suggest that the effective temperature of the star is 
significantly hotter than the value determined from the survey spectroscopy. 
However, the masses determined from these two models are not consistent with 
each other. More recently, a new asteroseismic analysis based on the full 
{\sl Kepler} data set provided by \citet{bi14} confirmed 
the former results found by \citet{bi11}.{We point 
out that a new asteroseismic analysis of KIC~08626021 is discussed} in 
\cite{gi15}.

KIC~08626021 has been observed by {\sl Kepler} for nearly two years in SC 
mode without interruption since the quarter Q10. Thus, it is a suitable candidate to 
investigate the long term amplitude and frequency modulations of the 
oscillation modes occurring in this star. In this paper we present a new 
thorough analysis of the {\sl Kepler} lightcurve obtained on the DB pulsator 
KIC~08626021, emphasizing in particular the time dependence of the 
amplitudes and frequencies of the modes associated to rotationally split 
triplets (Section 2). We provide {arguments} linking the uncovered 
amplitude 
and frequency modulations to the nonlinear mode coupling mechanisms 
(Section 3), before summarizing and concluding (Section 4).

\section{The frequency content of KIC 08626021 revisited}

The pulsating white dwarf star KIC 08626021 has been continuously observed 
by {\sl Kepler} in 
short cadence (SC) mode from quarter Q10.1 to Q17.2 (when the second 
inertial wheel of the satellite failed). A lightcurve from Q7.2, well 
disconnected from the main campaign, is also available for that star. 
Some analyses of these data have already been reported in the literature 
\citep{os11,co12}, including most recently the asteroseismic study of 
\citet[hereafter BK14]{bi14} based on the full Q10.1 -- Q17.2 data set. 
We initially considered using these published results as the starting point 
of our present study, but we realized that important details were lacking 
for our specific purposes. Consequently, we detail below, as a necessary 
step, our own thorough analysis of the frequency content of KIC~08626021.

\subsection{The {\sl Kepler} photometry}

All the data gathered by {\sl Kepler} for that star are now in the public domain. We
obtained the lightcurves from the Mikulski Archive for Space Telescopes 
(MAST)\footnote{https://archive.stsci.edu/}. As is standard, these data were processed 
through the {\sl Kepler} Science Processing Pipeline \citep{je10}. In the following, we concentrate 
on the consecutive data covering Q10.1 to Q17.2 without considering Q7.2 that would introduce 
a large time gap in the assembled lightcurve. With this restriction, we are left with a mere 
23 months of high precision photometric data starting from BJD 2 455 740 and ending on 
BJD 2 456 424 ($\sim$ 684 days) with a duty cycle of $\sim 87 \%$.

\begin{figure}
\includegraphics[width=8.5cm]{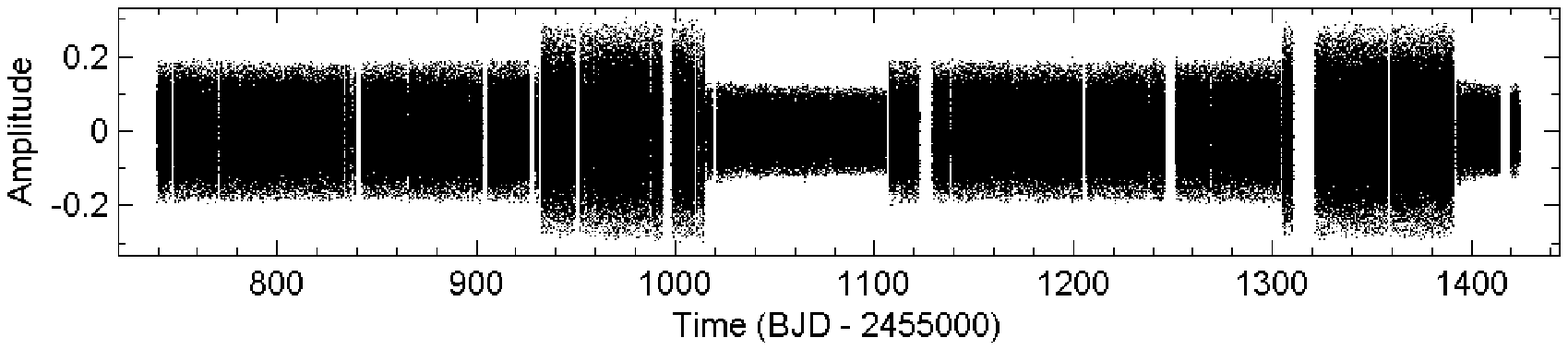}
\includegraphics[width=8.5cm]{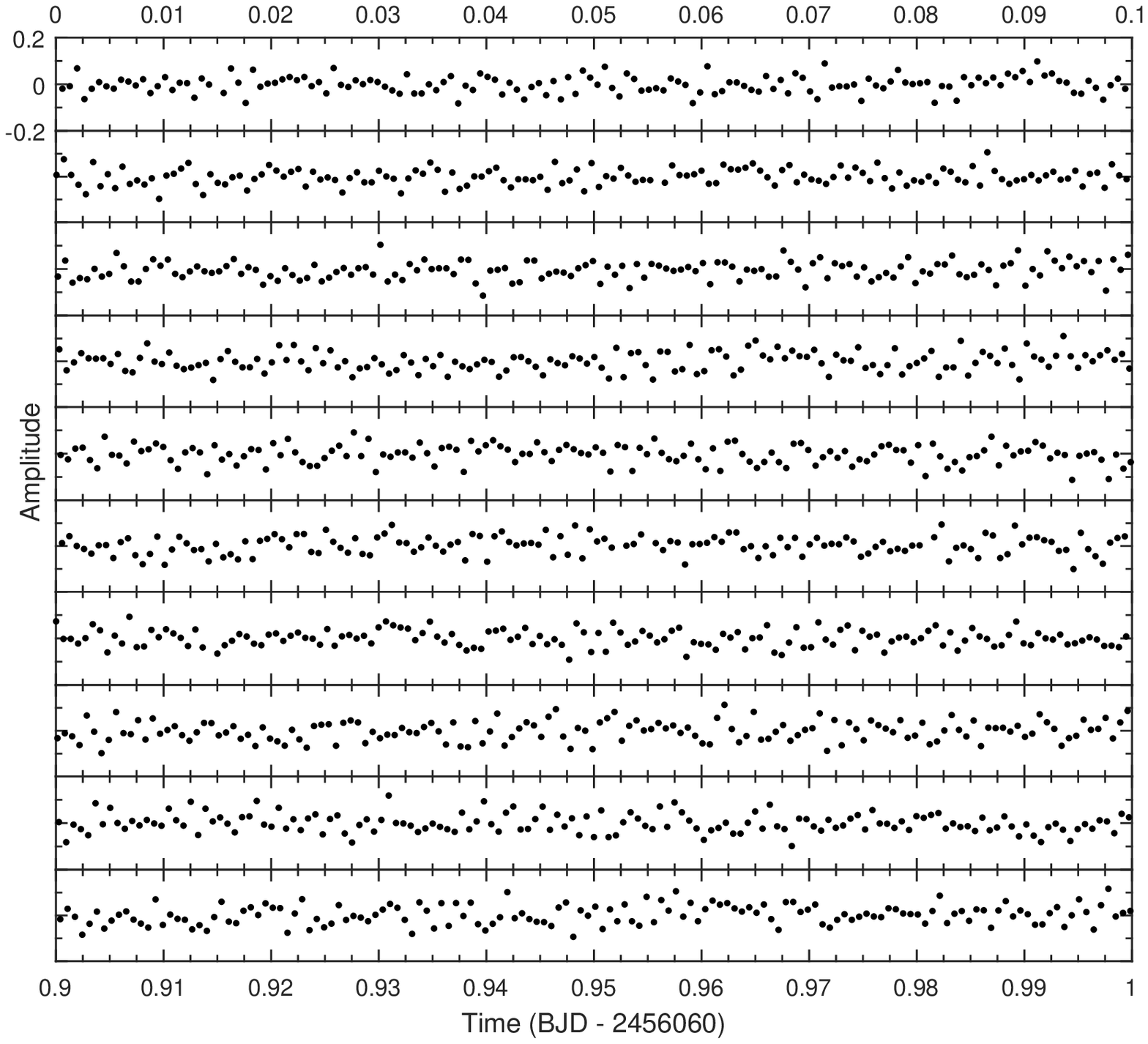}
\caption{{\sl Top panel}: Condensed representation of the full {\sl Kepler} lightcurve 
(Amplitude is the residual relative to the mean brightness 
intensity of the star vs time in Barycentric Julian Date)
covering from Q10.1 to Q17.2 ($\sim 684$ days) of KIC 08626021. 
Between quarters, scatter sometime changes due to different levels 
of noise in the data (all quarters are not equivalent in photometric 
quality). {\sl Bottom panel}: Close-up view showing one day of the {\sl Kepler} 
lightcurve by slices of 0.1 days. At this scale the oscillations are clearly apparent. 
\label{lc}}
\end{figure}

We constructed the full lightcurve from each quarter "corrected" lightcurves,
which most notably include a correction for the amplitude due to the contamination 
of the star by a closeby object (this correction consider that only $\sim 47.3\%$ of 
the light comes from the DB white dwarf). Tests indicate that the main 
differences between these corrected data and the raw data set used by BK14
occur in the measured amplitudes of the light variations, but has otherwise 
no noticeable incidence on the extracted frequencies. Each quarter light curve 
was then individually corrected to remove residual long term trends (using sixth-order 
polynomial fits) and data points differing significantly from the local standard deviation 
of the lightcurve were removed by applying a running 3-$\sigma$ clipping filter. 
The later operation {just} very slightly decreases the overall noise 
level.

\begin{figure}
\includegraphics[width=8.5cm]{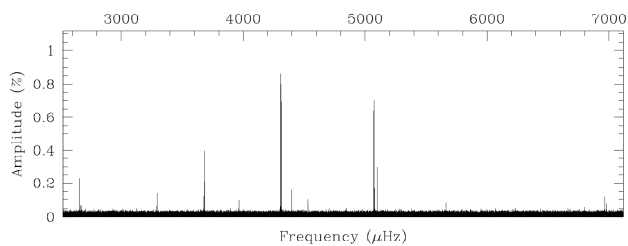}
\caption{Lomb-Scargle Periodogram (LSP; Amplitude in \% of the mean brightness
vs frequency in $\mu$Hz) of the {\sl Kepler} lightcurve. 
The represented range covers the frequency region where the 
pulsation modes are detected (no significant signal is found outside these limits).
Note that some well known {\sl Kepler} instrumental artefacts are within this 
frequency range but can easily be recognized.
\label{lsp}}
\end{figure}

The resulting ligthcurve is shown in Figure 1 and the corresponding Lomb-Scargle 
Periodogram (LSP; \citealt{sc82}) is given in Figure 2. The low-amplitude 
multi-periodic modulations are clearly seen with dominant periodicities of 
the order of a few minutes, typical of $g$-mode oscillations observed in 
pulsating DB white dwarfs. The formal frequency resolution in the Lomb-Scargle 
periodogram (defined as the inverse of the total time base line of the 
observations) reaches $\sim 0.017$ $\mu$Hz.

\subsection{Defining a secure detection threshold}

Before proceeding with the extraction of the frequencies, a brief discussion 
of the criteria used to define the confidence level of the detections is necessary.
With ground based observations of pulsating compact stars, a widely used rule of 
thumb was to consider the limit of $4\sigma$ (4 times the average local noise 
in the Fourier Transform) as the threshold above which a signal could safely be 
considered as real. However, with space observations, in particular with 
{\sl Kepler}, it became increasingly clear that this rule underestimates the 
risks of false detections resulting from statistical noise fluctuations. The 
reason lies most probably in the very large number of data points collected 
during months (or years) of observations with a sampling time of only 58s 
in SC mode. In particular, more than half a million frequency bins are necessary
to represent the Lomb-Scargle Periodogram of the 684 days {\sl Kepler} photometric 
data of KIC 08626021 and noise fluctuations are very likely to occur at least 
one time (and more) above a standard $4\sigma$ threshold. For this reason, the 
trend has been to increase the threshold to higher S/N values in somewhat 
arbitrary ways to avoid false detections (e.g., BK14 just assumes that 
the acceptable limit is $5\sigma$).

\begin{figure}
\includegraphics[width=8.5cm]{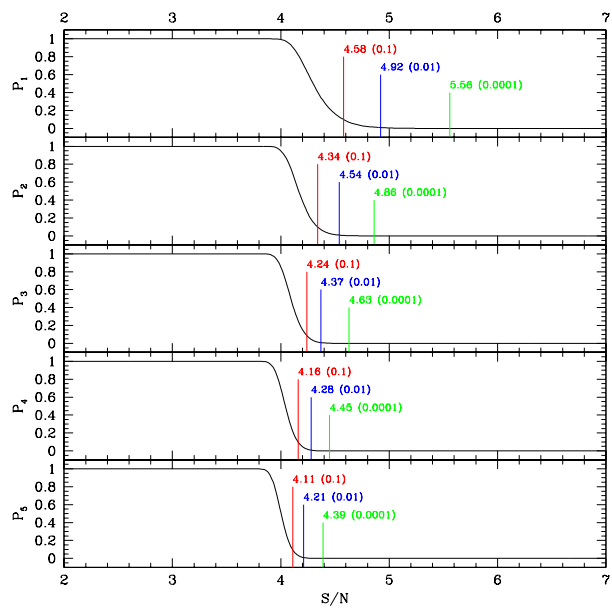}
\caption{False alarm probability $P_n$ of having at least $n$ peaks caused by 
noise  above a given signal-to-noise ratio (S/N) threshold in the Lomb-Scargle 
periodogram of KIC 08626021 (see text for details). In each panel, 
the 3 vertical lines indicate, from left to right, 
the S/N value corresponding to $P_n = 0.1$, 0.01, and 0.0001, respectively.
The limit $P_1 = 0.0001$, meaning that the risk of having a false detection 
above that threshold is reduced to less than 1 chance over 10 000, defines 
our adopted criterion. It corresponds to the limit of $5.6\sigma$. 
\label{sn}}
\end{figure}

Instead of adopting an arbitrary value, we quantitatively estimate what 
should be an acceptable threshold with the following procedure. Using the same
time sampling as the observations, we randomly build 10 000 artificial 
lightcurves just containing white gaussian noise (a random normal deviate 
is calculated at each time point). The Lomb-Scargle Periodograms of these 
artificial lightcurves are then calculated, as well as the median values of the noise 
in each resulting LSPs. For any given S/N threshold ($x\sigma$ defined 
as $x$ times the median noise level) we then find the number of times 
that at least $n$ peaks in the LSP (which by definition are just noise 
structures) happen to be above the chosen {limit}. Then, dividing by 
the number of tests (10 000 here), we obtain the false alarm probability 
$P_n(x)$ that at least $n$ peaks above a given S/N threshold of $x\sigma$
is due to noise.

\begin{figure}
\includegraphics[width=8.5cm]{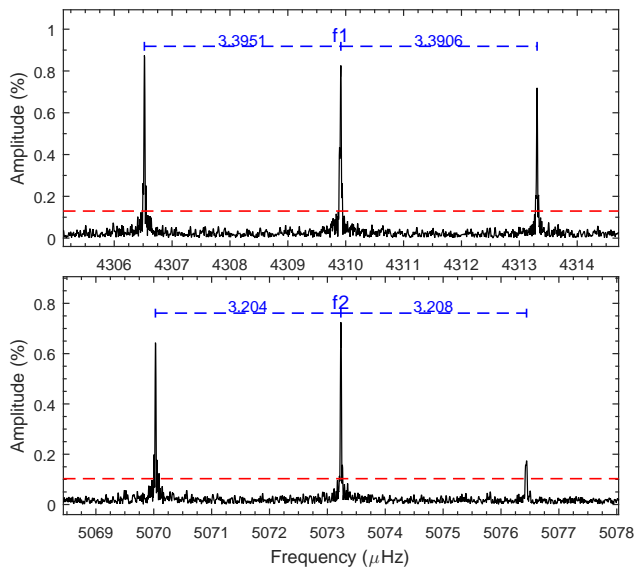}
\caption{Lomb-Scargle Periodogram of the triplets at 4310\,$\mu$Hz (up panel) 
and 5073\,$\mu$Hz (bottom panel) 
after nearly two years of {\sl Kepler} photometry gathered on KIC~08626021. 
The nearly equal frequency spacings are marked in the 
two panels. The dashed red horizontal line refers to the 5.6$\sigma$ detection 
threshold of local noise level. \label{triplets}}
\end{figure}

\begin{figure}
\includegraphics[width=8.5cm]{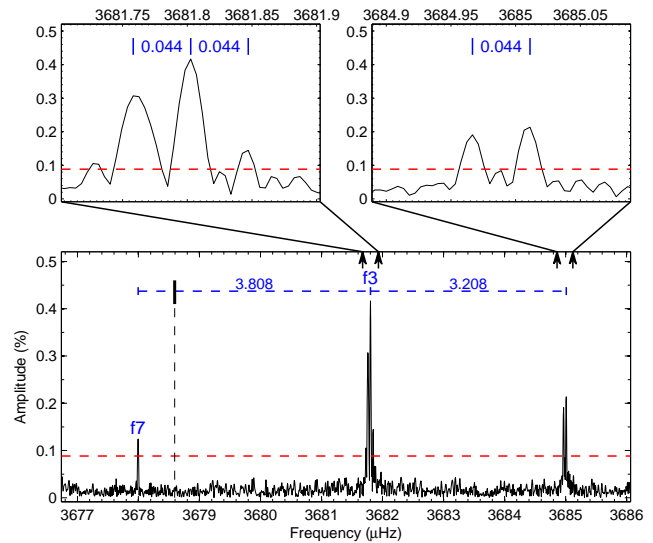}
\caption{Lomb-Scargle Periodogram in the 3677$-$3686\,$\mu$Hz frequency range where a doublet and a independent frequency are 
above the the 5.6$\sigma$ detection threshold of local noise level. The bold black vertical segment and vertical dashed line 
indicate the position of one possible third component for the doublet, as shown in the bottom panel. The two up panels are 
the expanded view 
of the two components as indicated by the vertical arrows in the bottom panel. Both the fine structures of the 
two components show well resolved closed peaks and share the same frequency separation of 0.044\,$\mu$Hz.
\label{doublet}}
\end{figure}

Figure~\ref{sn} shows the results of this procedure for the probabilities 
$P_1$ to $P_5$ as functions of the S/N threshold. The most interesting case 
is $P_1$ (the probability that at least 1 peak due to noise is above the 
threshold). We clearly see here that at the usual 4$\sigma$ limit, the 
probability to have at least one false detection is close to 1 (and $\sim 0.5$
to have at least 5 false detections according to $P_5$) confirming that 
this threshold is particularly unsafe in our case. However, $P_1$ eventually 
decreases with increasing S/N to reach 0.1 (10\% chance) at S/N $\sim 4.58$, 
0.01 (1\% chance) at S/N $\sim 4.92$ (approximately the detection threshold 
chosen by BK14), and less than 1 chance out of 10 000 at 
S/N = 5.56 (this is the limit above which not a single peak due to noise 
has been found among the 10 000 randomly generated lightcurves). 

Based on these calculations, we adopt in the following the conservative 
$5.6\sigma$ threshold as our limit of detection.

\begin{table*} \caption[]{List of frequencies detected in KIC 08626021.}
\small
\begin{center}
\begin{tabular}{ccccccccccl}  \hline \hline
Id.          &Frequency   &$\sigma_f$ & Period  &$\sigma_P$&Amplitude&$\sigma_A$&Phase&$\sigma_\mathrm{Ph}$&S/N &Comment  \\
&             ($\mu$Hz)   &($\mu$Hz)  &(s)      &(s)       &(\%)     &(\%)                   \\
\hline
&&\\
$f_{1,-}$    & 4306.52304 & 0.00013& 232.205886 & 0.000007 & 0.867 & 0.012 & 0.7987 & 0.0037  & 73.4 & $f_{1,-}$ in BK14\\
$f_{1,0}$    & 4309.91490 & 0.00014& 232.023143 & 0.000007 & 0.804 & 0.012 & 0.5264 & 0.0040  & 68.1 & $f_{1,0}$ in BK14\\
$f_{1,+}$    & 4313.30642 & 0.00016& 231.840705 & 0.000008 & 0.701 & 0.012 & 0.7885 & 0.0046  & 59.3 & $f_{1,+}$ in BK14 \\
&&\\
$f_{2,-}$    & 5070.03081 & 0.00017& 197.237460 & 0.000007 & 0.641 & 0.012 & 0.1521 & 0.0050  & 54.3 & $f_{2,-}$ in BK14\\
$f_{2,0}$    & 5073.23411 & 0.00016& 197.112922 & 0.000006 & 0.705 & 0.012 & 0.0394 & 0.0046  & 59.8 & $f_{2,0}$ in BK14\\
$f_{2,+}$    & 5076.44385 & 0.00066& 196.988291 & 0.000026 & 0.167 & 0.012 & 0.1462 & 0.0192  & 14.1 & $f_{2,+}$ in BK14\\
&&\\
$f_{3,0}$\tablefootmark{$\dagger$} & 3681.80286 & 0.00028& 271.606068 & 0.000020 & 0.397 & 0.012 & 0.1347 & 0.0082  & 33.6 & $f_{3,0}$ in BK14\\
$f_{3,+}$\tablefootmark{$\dagger$} & 3685.00937 & 0.00052& 271.369731 & 0.000038 & 0.212 & 0.012 & 0.4066 & 0.0153  & 18.0 & $f_{3,+}$ in BK14\\
&&\\
$f_4$        &2658.77740  & 0.00047& 376.112721 & 0.000067 & 0.233 & 0.012 & 0.6147 & 0.0140  & 19.7 & $f_{5}$ in BK14\\
$f_{5}$      & 4398.37230 & 0.00068& 227.356834 & 0.000035 & 0.161 & 0.012 & 0.7598 & 0.0200  & 13.6 & $f_{7}$ in BK14\\
$f_{6}$      &3294.36928  & 0.00079& 303.548241 & 0.000073 & 0.139 & 0.012 & 0.0934 & 0.0234  & 11.8 & $f_{4,0}$ in BK14\\ 
$f_{7}$      & 3677.99373 & 0.00088& 271.887358 & 0.000065 & 0.125 & 0.012 & 0.6773 & 0.0260  & 10.6 & $f_{3,-}$ in BK14\\
$f_9$        & 6981.26129 & 0.00139& 143.240592 & 0.000028 & 0.079 & 0.012 & 0.0105 & 0.0404  & 6.7 & $f_{11}$ in BK14\\
&&\\
\multicolumn{3}{l}{Linear combination frequencies}\\
$f_8$        & 6965.30234 & 0.00090& 143.568786 & 0.000019 & 0.121 & 0.012 & 0.8358 & 0.0264  & 10.3 & $f_{1,-}+f_4$ ; $f_{6}$ in O13\tablefootmark{a}\\
$f_{10}$     &2667.95462  & 0.00164& 374.818969 & 0.000230 & 0.067 & 0.012 & 0.7489 & 0.0484  &5.7   & $f_{9}-f_{1,+}$ \\ 
&&\\
\multicolumn{3}{l}{Frequencies above 5\,$\sigma$ detection}\\
$f_{11}$\tablefootmark{$\star$}    &2676.38212 & 0.00170& 373.638725 & 0.000236 & 0.065 & 0.012 & 0.1443 & 0.0501 &5.5 \\
$f_{12}$\tablefootmark{$\star$}    &3290.24565 & 0.00176& 303.928675 & 0.000163 & 0.063 & 0.012 & 0.0752 & 0.0519 &5.3 & $f_{4,-}$ in BK14\\
\hline
\end{tabular}
\end{center}
\label{t1}
\tablefoot{
\tablefoottext{$\dagger$}{The azimuthal order ($m$) identification proposed for the doublet comes from a model fit seismic 
solution (see Giammichele et al. 2015);}  
\tablefoottext{$\star$}{These frequencies are below $5.6\sigma$ (although still higher than $5\sigma$) and could be spurious;} 
\tablefoottext{a}{\citet{os13}.}
}
\end{table*}

\subsection{Extraction of the frequencies}

We used a dedicated software, \felix{} (Frequency Extraction for LIghtcurve 
eXploitation) developed by one of us (S.C.), to first extract the frequency content
of KIC 08626021 down to our adopted detection threshold of $5.6\sigma$ (we, 
in practice, pushed down the limit to $\sim 5\sigma$; see below). The method 
used is based on the standard prewhithening and nonlinear least square fitting 
techniques \citep{de75} that works with no difficulty in 
the present case. The code \felix{} greatly eases and accelerates the application of 
this procedure, especially for long and consecutive time series photometry 
obtained from spacecrafts like CoRoT and {\sl Kepler} \citep{ch10,ch11}.

\begin{figure}
\includegraphics[width=8.5cm]{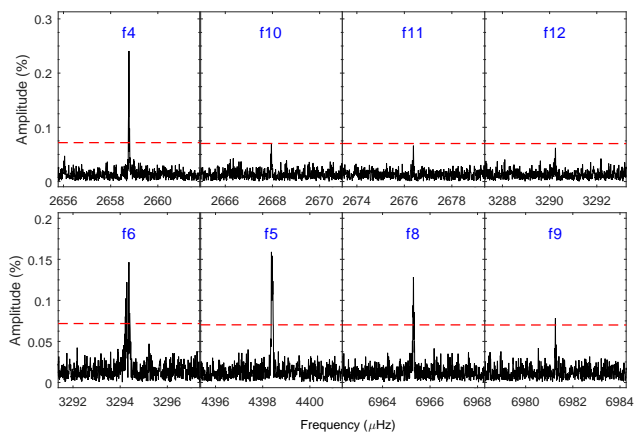}
\caption{Lomb-Scargle Periodogram of 8 frequencies after nearly two years of {\sl Kepler}
        photometry gathered on KIC~08626021. The two frequencies $f_{11}$ and $f_{12}$ are with 
        5.0$-$5.5\,$\sigma$ of local noise level. Frequency $f_{8}$ and $f_{10}$ are 
        linear combination frequencies.
        \label{otherfreqs}}
\end{figure}

The list of extracted periodic signals is provided in Table~\ref{t1} which 
gives their fitted attributes (frequency in $\mu$Hz, period in seconds, 
amplitude in \% of the mean brightness, phase relative to a reference 
$t_0$ time, and signal-to-noise above the local median noise level) along 
with their respective error estimates ($\sigma_f$, $\sigma_P$, $\sigma_A$, 
and $\sigma_\mathrm{Ph}$). Figure~\ref{triplets}, \ref{doublet}, and 
\ref{otherfreqs} show zoomed-in views of all the identified peaks in 
the Lomb-Scargle periodogram. 

We find 13 very clear independent frequencies that come out well above the 
detection threshold. Two additional lower amplitude peaks ($f_8$ and $f_{10}$) 
appear as significant but are linked to other frequencies through linear 
combinations and are therefore likely not independent pulsation modes. Two more 
frequencies ($f_{11}$ and $f_{12}$) can be identified above $5\sigma$ but 
below $5.6\sigma$ which we mention for completeness, but that cannot be 
considered as secured detections. A comparison with the completely independent 
analysis of BK14 shows that we agree on all the relevant, well secured 
frequencies (i.e, with a sufficiently high S/N ratio). We point out however 
that some additional features of the frequency spectrum are not discussed in 
BK14 and we differ on how to interpret some of the mode associations 
(see below).

As reported in BK14, six of the extracted frequencies ($f_1$ and $f_2$) 
form 2 very well defined, nearly symmetric triplets with a frequency spacing of 
$\sim 3.392$ $\mu$Hz and $\sim 3.206$ $\mu$Hz (Fig.~\ref{triplets}). These are 
readily interpreted as rotationally split triplets, thus giving an average 
rotation period of $\sim 1.75$ day {for the star}. However, we argue 
that the 3 frequencies
shown in Figure~\ref{doublet} cannot correspond to the components of a 
triplet, as BK14 suggest. These frequencies form a clearly asymmetric
structure with the left component ($f_7$) being significantly more distant 
than the right component ($f_{3,+}$) from the central peak ($f_{3,0}$). 
We note in this context that the frequency separation between $f_{3,0}$ 
and $f_{3,+}$ ($3.208$ $\mu$Hz) is similar or very close to the frequency 
splitting characterizing the $f_2$ and $f_1$ triplets. Our interpretation 
is therefore that the middle ($f_{3,0}$) and right ($f_{3,+}$) peaks 
are 2 components of a triplet whose 3rd component is undetected, while 
the left peak ($f_7$) is a completely independent mode. This has some 
implications in finding an asteroseismic solution for KIC 08626021 as 
attempted by BK14, since 8 independent periods should be considered and 
not 7 (see \citealt{gi15}). 

Furthermore, not reported in BK14, we show in Figure~\ref{doublet} 
that the two components of the incomplete triplet are in fact surrounded 
by additional structures (not tabulated in Table~\ref{t1}). The central 
peak ($f_{3,0}$) appears to have 2 resolved symmetric sidelobes located 
$0.044$ $\mu$Hz away, while the right peak ($f_{3,+}$) shows a sidelobe 
also separated by $0.044$ $\mu$Hz. These intriguing "hyperfine" 
structures cannot be associated to rotation since a much larger 
rotational splitting signature has already been found. Moreover, the very small 
frequency separation involved would indicate a modulating phenomenon 
that occurs on a very long timescale of $\sim 263$ days.

This finding brings us to the main {subject} of the present 
paper, which is to {show} that this "hyperfine" structure, along 
with other behaviors that we discuss below, {can be linked} to long 
term amplitude and frequency modulations generated by nonlinear resonant 
coupling mechanisms between the components of rotationally split triplets.

\subsection{Amplitude and frequency modulations}

From now on, we mainly focus our discussions on the two well defined 
triplets $f_1$ and $f_2$, and on the "doublet" $f_3$ (i.e., the two visible 
components of an incomplete triplet). In order to analyze the time variability
of these modes and their relationship, we used our software \felix{} to compute 
the sliding Lomb-Scargle periodogram (sLSP) of the data set. This technique 
consists of building time-frequency diagrams by filtering in only parts of the data
as a function of time. In the present case, we chose a filter window that is 180 day 
wide sled along the entire lightcurve by time steps of 7 days. This ensures 
a good compromise, for our purposes, between time resolution, frequency 
resolution (to resolve close structures in each LSP), and signal-to-noise. 
The sLSP gives an overall view of the amplitude and frequency variability that
could occur for a given mode (see, e.g., the top left panel of 
Fig.\,\ref{pf_mod}). We acknowledge that BK14 also provide a similar 
analysis, but they chose a sliding window that is only 14 day wide, hence 
providing a much lower resolution in frequency. This has strong consequences 
on the interpretation of these data that will become obvious below. 
As a complementary (and more precise) method, we also extracted the frequencies 
(through prewithening and nonlinear least square fitting techniques) in
various parts of the lightcurve, i.e., the 23-month light curve of 
KIC 08626021 was divided into 20 time intervals, each containing 6 months 
of data (for precision in the measurements) except for the last 3 intervals 
at the end of the observations. This second approach provides a measure of 
the (averaged) frequencies and amplitudes at a given time, along with the 
associated errors (see, e.g.,  the bottom left panel of Fig.\,\ref{pf_mod}).

\subsubsection{The $f_1$ triplet}

\begin{figure*}
\centering
\includegraphics[width=15.5cm]{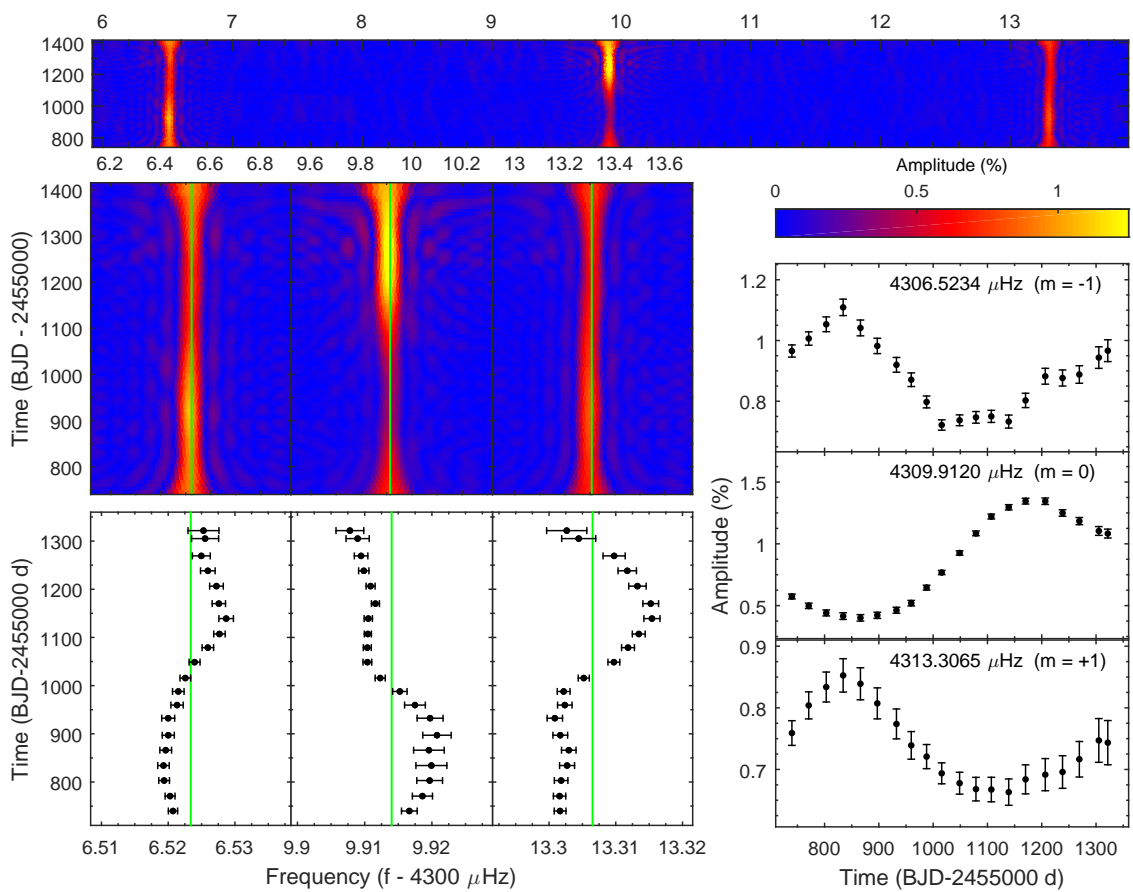}
\caption{Frequency and amplitude modulations in the $f_1$ triplet at 
4310\,$\mu$Hz. The top panel shows the sliding Lomb-Scargle Periodogram 
(sLSP giving the amplitude in \% as a function of frequency in $\mu$Hz and time 
in days) of the triplet as a whole. The middle and bottom left panels show 
increasingly expanded views around the average frequency (the solid vertical 
lines) of each component, with the bottom left panel obtained from 
prewhitening subsets of the data, thus measuring precisely the frequencies, 
as a function of time. The lower right panel provides the measured amplitudes 
as a function of time obtained for each subset of data 
(see text for details).\label{pf_mod}}
\end{figure*}

Figure\,\ref{pf_mod} shows the amplitude and frequency modulations 
observed 
for the 3 components forming the $f_1$ triplet near 4310 $\mu$Hz. In this 
plot, views of the frequency variations with time are illustrated from top 
to bottom-left panels. The top panel first shows the sLSP of the triplet 
as a whole (similar to Figure 2 of BK14) where the signal appears, 
at this scale, stable in frequency but varying in amplitude for at least 
the central component. Then we provide increasingly expanded views (from 
middle-left to bottom-left panel) around the average frequency of each 
component. In addition, the bottom right panel shows how the amplitude of 
each component varies with time.

\begin{figure}
\centering
\includegraphics[width=8.5cm]{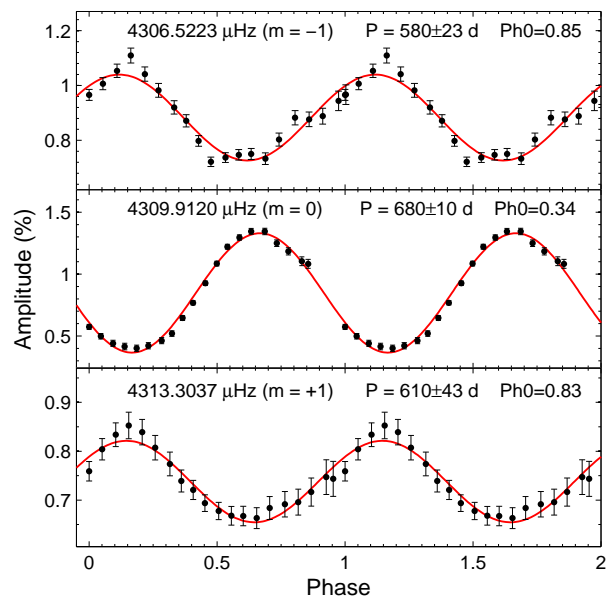}
\caption{{Amplitude modulations observed in the $f_1$ 
triplet presented in phase diagrams with a folding period of 580 and 610  
days for the two side components, and 680 days for the central component. 
These periods, and their associated formal errors, were estimated 
by fitting a sinewave (red curve) to the amplitude variations.}
\label{ammodu_repeat}}
\end{figure}

\begin{figure}
\centering
\includegraphics[width=7.5cm]{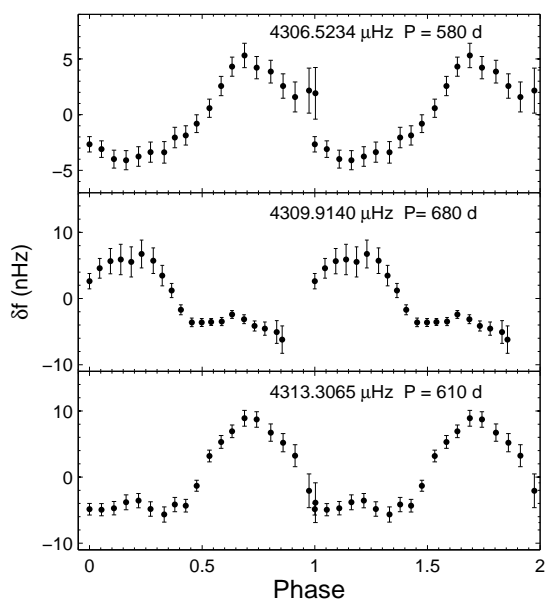}
\caption{ Same as Fig.~\ref{ammodu_repeat} but for the frequency
modulations observed in the $f_1$ triplet. Note that the folding periods
adopted are the values derived for the corresponding amplitude modulations.
\label{frmodu_repeat}}
\end{figure}

\begin{figure*}
\begin{center}
\includegraphics[width=15.5cm]{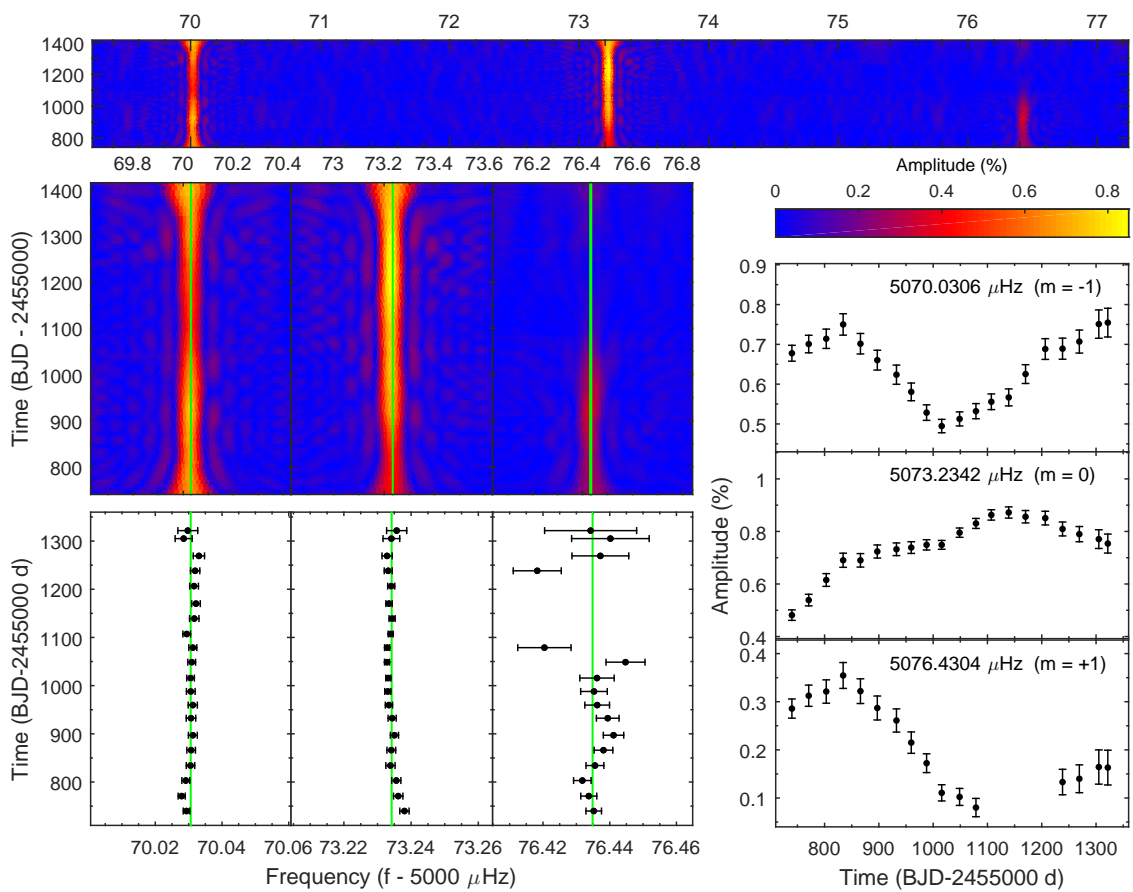}
\end{center}
\caption{Same as Figure\,\ref{pf_mod} 
but for the $f_2$ triplet at 5073\,$\mu$Hz. Note that the component
at 5076\,$\mu$Hz has a very low amplitude and plunges below the
4\,$\sigma$ detection threshold during the second half of the 
observations, making it difficult to follow (hence the 4 missing 
data points).\label{st_mod}}
\end{figure*}

It is mentionned in BK14 that the modes, and these 3 components 
in particular, are stable in frequency over the 2-year duration of the 
observations. We clearly demonstrate here that this is not the case. Their 
statement is based on a time-frequency analysis involving a sliding Fourier 
Transform (sFT) that only uses a 14d-wide window, which clearly does not 
permit a sufficient frequency resolution to uncover the modulations that 
we report here. We find that both the amplitudes and frequencies show
{very suggestive} signatures of quasi periodic modulations with an 
average timescale that we can roughly estimate to $\sim620$ days. 
{Figure\,\ref{ammodu_repeat}
and \ref{frmodu_repeat} illustrate further this periodicity in phase 
diagrams.
Although very similar, we find that the modulation period associated to 
the side components of the triplets ($\sim 600$ d) could be slightly 
shorter than the modulation period of the central component ($\sim 
680$ d). Note that the modulating periods in this triplet were obtained by 
searching for the best fit of a pure sine wave function to the 
amplitude variations. The zero phase is relative to the time of 
the first data point (BJD=2\,455\,739.836). 
The same procedure was not applied to the frequency modulations 
since the patterns are clearly more complex than a pure sine wave 
function. However, since Fig.~\ref{pf_mod} suggests a cyclic behavior
with roughly the same timescale, the folding periods used to construct
Fig.~\ref{frmodu_repeat} were chosen to be those derived for the amplitude 
modulations. This allows us to check that indeed at least two of the 
components (the $m=+1$ and $m=-1$ modes) accomodate rather well these 
periodicities, as the curves connect near phase 1\footnote{Note that the 
two last data points with larger error bars are less secure measurements 
due to the fact that the sliding Fourier transform reaches the end of 
the data set and incorporates shorter portions of the lightcurve. This 
affects the frequency resolution and consequently the precision of the 
measurements.}. For the central ($m=0$) component, a slightly longer 
estimated periodicity does not permit to cover entirely the suspected 
modulation cycle with the data, leaving a gap between phase 0.9 
and 1 where the behavior is not monitored. We cannot say in that case 
whether this curve would eventually connect smoothly at phase 1 or if 
a discontinuity exists, suggesting that either the chosen 
folding period is not appropriate or an additional trend is affecting the 
frequency of this mode.  

In addition, we note that the frequency and amplitude modulations show 
obvious correlations, as both evolve in phase with the same period 
(with period of $\sim600$ days and zero phase of $\sim0.84$),
for the side components, and are somewhat antiphased with the central component
(with zero phase of $0.34$),
as shown in Fig.\,\ref{ammodu_repeat} and \ref{frmodu_repeat}. 
{We quantitatively checked this fact by computing the 
correlation coefficients between, e.g., the amplitudes of the $m=+1$ and 
$m=-1$ components ($\rho_{A_+,A_-} = +0.93$; i.e., indicative of a strong 
correlation) and the amplitudes of the $m=+1$ and 
$m=0$ components ($\rho_{A_+,A_0} = -0.82$; i.e., indicative of a strong 
anti-correlation). 
Such correlated behavior suggests that the modes involved are somehow 
connected, either through a common cause affecting similarly their amplitude 
and frequency or through direct interactions occurring between the 
components of the triplet. This will be discussed further in Section 3.

\subsubsection{The $f_2$ triplet}

\begin{figure*}
\begin{center}
\includegraphics[width=15.5cm]{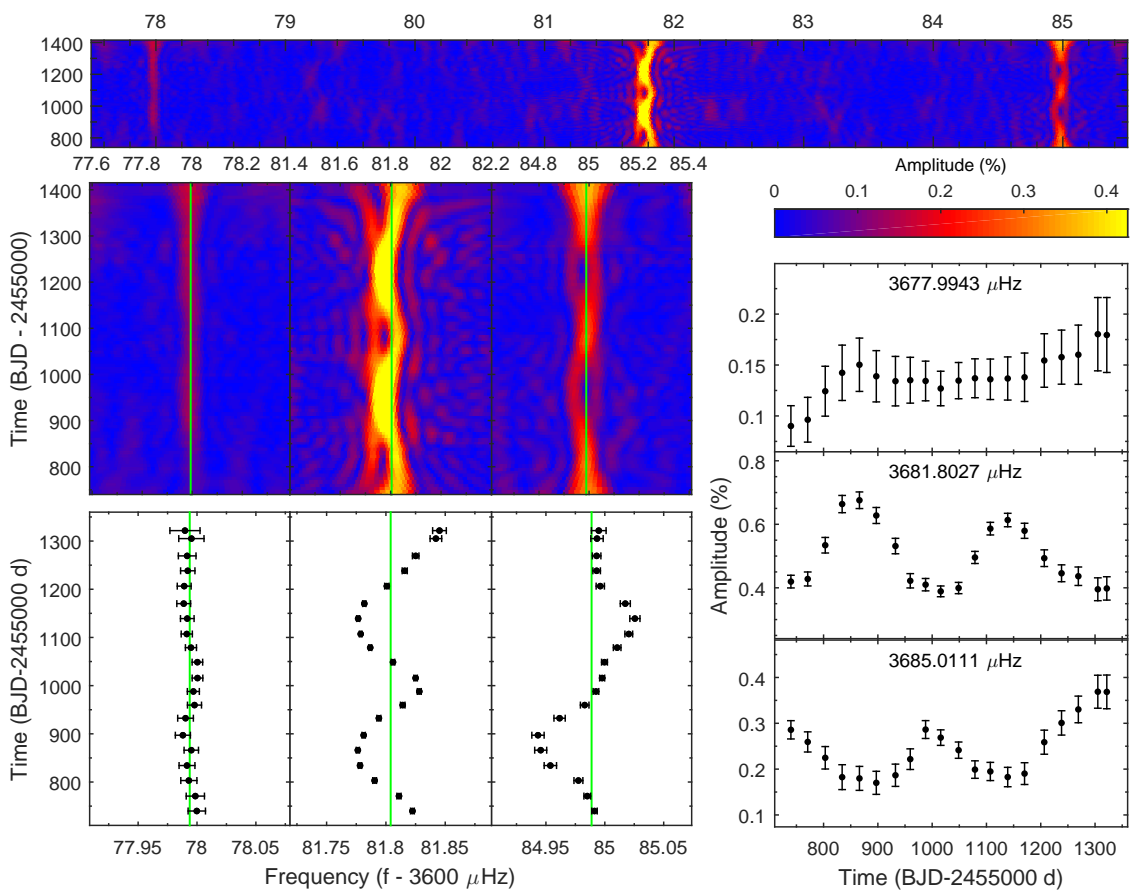}
\end{center}
\caption{Same as Figure\,\ref{pf_mod} 
but for the doublet $f_3$ at 3682\,$\mu$Hz and the nearby independent 
frequency $f_7$ at 3678\,$\mu$Hz.\label{tt_mod}}
\end{figure*}

Figure\,\ref{st_mod} shows the modulations observed in the other triplet, $f_2$, at 
5073\,$\mu$Hz. The frequencies in this triplet appear to be stable during the 
nearly two years of {\sl Kepler} observations, while the amplitudes show clear 
modulations. Note that the amplitude of the $m=+1$ component went 
down at some point below a signal-to-noise ratio of $4\sigma$ and was 
essentially lost in the noise during a portion of the last half of the 
observations. Four measurements could not be obtained because of this and 
when it was still possible to spot this component, the errors remained large.

{Again in this case, the amplitudes of the two side components seem to 
evolve nearly in phase with a quasi-periodic behavior on a timescale that 
is probably slightly larger than the duration of the observations and 
close to a timescale of $\sim 700$ day}. However, contrary to the previous 
case, a connection with the central component is less clear. The later 
seems to follow a variation pattern possibly occurring on a longer time 
scale. {Therefore, $f_2$ behaves somewhat differently 
from $f_1$, a feature that we will discuss more in the next section}.

\subsubsection{The $f_3$ doublet}

The 3677$-$3686\,$\mu$Hz frequency range is shown in Fig.\,\ref{tt_mod} and 
contains the independent frequency ($f_7$) and the two visible components 
of the incomplete triplet $f_3$ (thus forming a doublet). 

Each component of this doublet shows clear signatures of correlated 
variations for both amplitudes and frequencies. 
We note in particular a periodic modulation that occurs on a somewhat 
shorter timescale than for the two previous cases. 
A very quick look at Fig.\,\ref{tt_mod} indicates a period of 
roughly 280 days for the amplitude variations of both modes as well as 
for the frequency modulation of $f_{3,0}$, which in fact can readily be 
connected to the "hyperfine" structure sidelobes discussed in Section 2.3 and 
illustrated in Fig.~\ref{doublet}. The frequency of $f_{3,+}$,
for its part, seems to also follow a periodic trend but, quite interestingly, 
on a timescale that could be around twice ($\sim 550$ days) the period of 
the other components.

It appears now clearly that a periodic frequency and amplitude modulation 
process is responsible for the equidistant peaks surrounding $f_{3,0}$ 
and $f_{3,+}$. In this context, the 0.044 $\mu$Hz frequency separation 
should provide a more precise estimate of the period of this modulation, which 
is $\sim 263$ days. Note that the two other triplets discussed previously 
do not show similar "hyperfine" splitting structures around their components 
simply because the period of their modulations appear to be slightly longer 
than the observational time baseline and those structures cannot be resolved.
In the case of $f_3$, the observations are long enough to resolve the 
modulation. We further note that the amplitudes of the two components of 
$f_3$ evolves in antiphase while the frequencies are in phase during the first 
half of the run but then evolve in antiphase during the last part 
of the observations, which reflects the fact that the frequency 
variation of $f_{3,+}$ has approximately twice the period of the modulation 
seen in the frequency of $f_{3,0}$.

In contrast, the mode $f_7$ shows a totally different behavior as 
both its frequency and amplitude appear stable throughout the 
observing run. This could further support, if need be, the interpretation 
that $f_7$ and the $f_3$ complex are not part of a same triplet structure 
(as assumed by BK14). We indeed note that the theoretical framework in 
which these modulations can possibly be understood (nonlinear resonant 
couplings, as discussed in Section 3) forbids the possibility that the 
components of a triplet behave in different regimes.

\begin{figure*}
\begin{center}
\includegraphics[width=15.5cm]{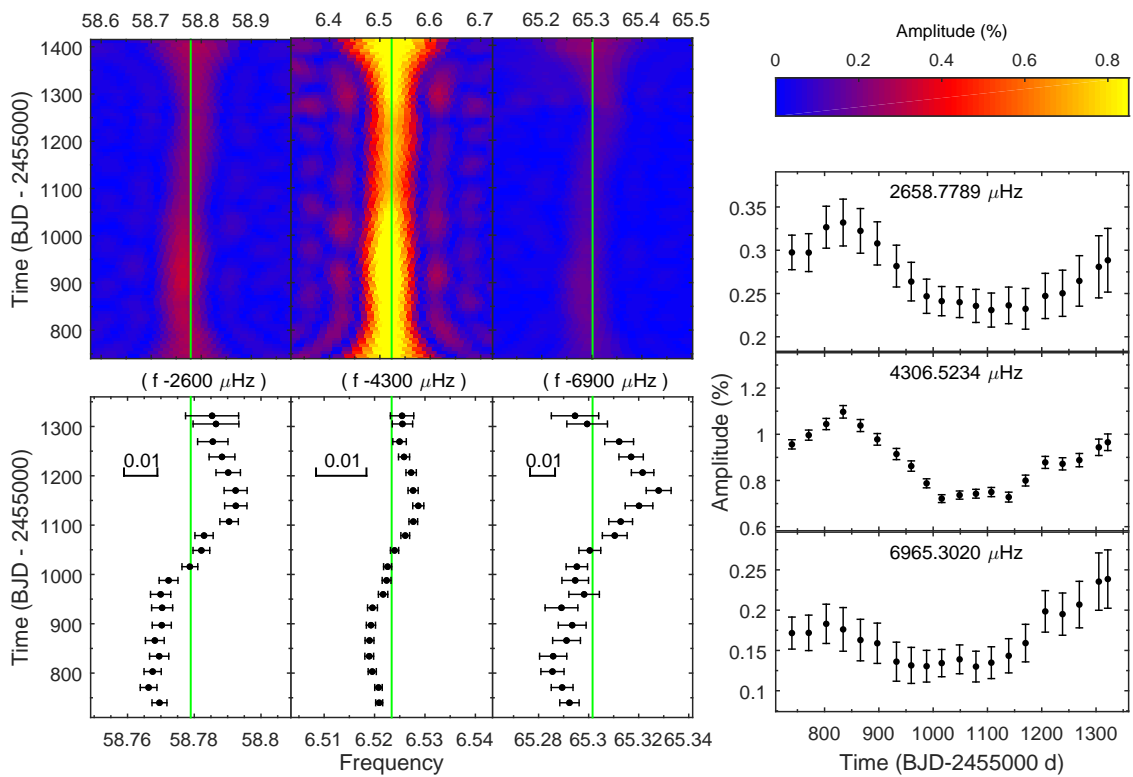}
\end{center}
\caption{Frequency and amplitude modulations of the components forming the 
linear combination $f_{1,-}+f_4 = f_8$. 
The panels are similar to those described in Figure\,\ref{pf_mod}.
The three components show clear periodic modulations evolving in phase. 
In each bottom left subpanels, a segment indicates the 0.01\,$\mu$Hz 
scale.  
\label{c_mod}}
\end{figure*}

\subsubsection{Other correlated modulations}

For completeness, we also illustrate the interesting behavior of 3 frequencies 
related by a linear combination. Figure~\ref{c_mod} shows the amplitude and frequency 
modulations of $f_{1,-}$, $f_4$ and $f_8$, that satisfy almost exactly 
(within $2\sigma$ of the formal measurement errors) the relation $f_8 = f_{1,-} + f_4$ 
(see Table~\ref{t1}).

It is striking to see how the 3 components follow nearly exactly the same 
trends in both frequency and amplitude. 
These modulations could be related to the so called parents/child mode 
nonlinear interactions discussed by \citet{br14} or to other 
nonlinearities encountered in white dwarfs (e.g., \citealt{br95,wu01a}).
In this context, we note, again according to the values given 
in Table~\ref{t1}, that the mean relative amplitude of $f_8$ ($A_8=0.00121$) 
is $\sim 60$ times larger than the product of the relative amplitudes of 
$f_{1,-}$ ($A_{1,-} =0.00867$) and $f_4$ ($A_4=0.00233$) whose value is 
$0.000020$. Figure\,\ref{lcf} shows that these relationships also hold 
(within $2\sigma$), both for frequencies and amplitudes, for each individual 
measurement done as a function of time illustrated in Figure~\ref{c_mod}. 
Interestingly, if this combination were to be related to the mechanism 
of \citet{wu01a}, the factor of $60$ connecting $A_8$ to $A_{1,-}.A_{4}$
would possibly imply that the inclination angle of the star should be 
$50^{\circ}<i<60^{\circ}$ (following Eqn.\,(20) in \citealt{wu01a}).
Alternately, $f_8$ could result from a resonant mode 
coupling phenomenon where $f_8$ is a true eigenmode of the star 
(possibly of higher degree $\ell$) whose amplitude is boosted above the 
detection limit by the resonance following Eqn.~(5) of 
\citet[see also \citealt{dz82}]{br14}. We indeed find that 
our results, instead of using phase (we here use frequency), 
are similar to the linear combination frequency families described in 
\citet[e.g., comparing their Fig.\,4-5 to our Fig.\,\ref{lcf}]{br14}. 
However, at this stage, we cannot decipher which of these potential 
mechanisms could explain the details of this combination of frequencies 
due to the lack of further independent constraints (such as the inclination 
angle of KIC\,08626021).
Finally, we point out that one of the frequencies involved in this relation, 
$f_{1,-}$, is also involved as one of the components of the $f_1$ triplet 
discussed in Section 2.4.1 (and illustrated in Fig.~\ref{pf_mod}). 

Another similar linear combination has also been identified, involving 
$f_{1,+}$, $f_9$ and $f_{10}$, but the low amplitudes of $f_9$ and $f_{10}$ 
have prevented us from analyzing its frequency and amplitude modulations. 
In the following, we concentrate on a possible theoretical 
interpretation of the frequency and amplitude modulations observed in the 
triplets, and we do not discuss further the properties of linear 
combination frequencies.

\begin{figure}
\centering
\includegraphics[width=8.5cm]{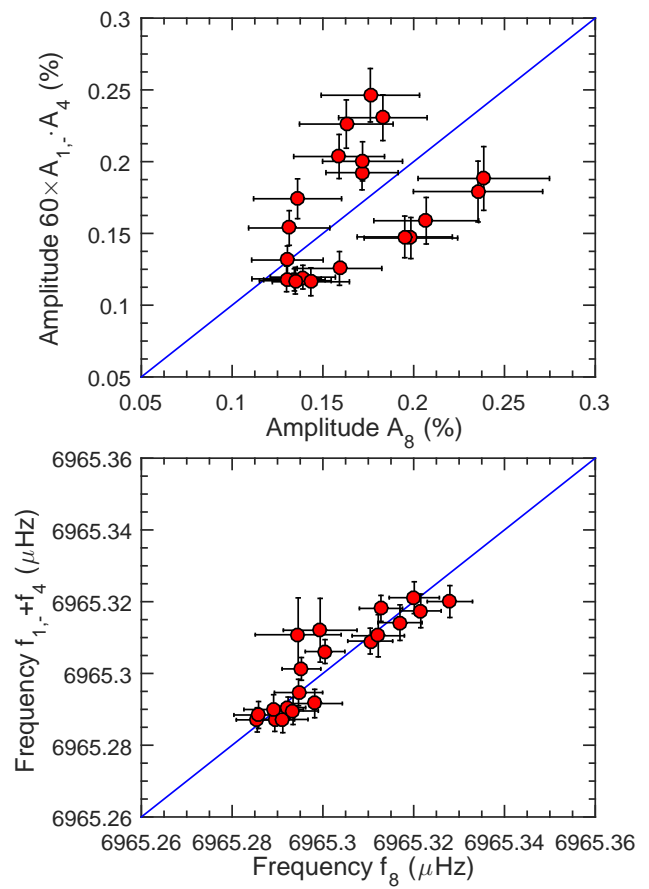}
\caption{{\sl Top panel}: Comparison between the measured amplitude 
$A_8$ of $f_8$ and 60\,$\times$\,$A_{1,-}\cdot{}A_4$ (where $A_{1,-}$ and 
$A_4$ are the amplitudes of $f_{1,-}$ and $f_4$, respectively) at each time 
segment (see text).
{\sl Bottom panel}: Same as top panel but for a comparison between the 
measured frequency $f_8$ and sum of the frequencies $f_{1,-}+f_4$. 
In both cases, the relationship is exactly statisfied along the plain 
diagonal line. All the measurements are found to be within $2\sigma$ of 
this line. 
\label{lcf}}
\end{figure}

\section{Links with nonlinear resonant couplings}

The frequency and amplitude modulations observed in the two triplets and the doublet 
of KIC 08626021 cannot be related to any evolutionary effect, such as neutrino cooling, 
because the timescale involved is several orders of magnitude shorter than the cooling 
rate of DB white dwarfs \citep{wi04}. The signature of orbiting companions around the 
star is also ruled out by the fact that the variations occurring in different 
frequencies are not correlated in phase and do not have the same amplitude 
modulations \citep{si07}. We also considered possibilities that instrumental 
modulations could occur, e.g., on a per quarter basis, such as a slightly 
varying contamination from the nearby star that could modulate the amplitude 
of the modes, but then all modes should be affected similarly, which is not 
what is observed. Finally, the possibility was raised that changes in the background state of 
the star, such as those induced by magnetic cycles or through an hypothetical 
angular momentum redistribution mechanism, could be responsible for the 
observed modulations. It is indeed well known that 
magnetic cycles have an impact on the frequencies of the $p$-modes observed in 
the Sun, leading to small frequency drifts that correlate well in time with 
tracers of the solar surface activity (see, e.g., \citealt{sa15} and 
references therein). One could imagine that such mechanisms may exist in 
white dwarfs as well. 
We find, however, that such effects would be hardly compatible with how the 
modes in KIC 08626021 are found to vary. In the Sun, all the modes appear to 
be globally affected following the same trends to some various extent, 
while in our case we see for instance a triplet $f_1$ that shows correlated 
changes in frequency, and at the same time another triplet $f_2$ whose 
frequencies appear to be constant. We find a mode $f_7$ that also does not 
change while the 2 visible components of the doublet $f_3$ nearby show 
correlated variations in frequency. This makes it difficult to connect these 
behaviors to a common global cause (i.e., small changes of the stellar 
structure). A cyclic redistribution of angular momentum, for its part, would 
affect the frequencies of the $m=+1$ and $m=-1$ components 
with an anti-correlation, while the central $m=0$ component should not be 
affected (and it is found to vary in $f_1$). All triplets should be 
affected nearly the same way, but $f_2$, showing constant frequencies, 
clearly is not and somewhat rules out this possibility.

Instead, we prefer to fall back to a simpler possibility. We develop 
in this section arguments that nonlinear resonant mode coupling mechanisms, 
by which both the amplitudes and frequencies of oscillation modes can be 
modulated on timescales of weeks, months, and even years, appears as 
a natural explanation for some of the observed behaviors.

\subsection{The amplitude equations formalism}

The amplitude equations (AEs) formalism {is, to our knowledge, the 
only existing theoretical tool to investigate mode couplings 
for nonradial oscillation modes in pulsating stars. AEs in the stellar  
context have been extensively studied since the 1980's for different types 
of couplings \citep{dz82,bu84,mo85,dz92,go94,van94,bu95,go98,wu01a,wu01b}. 

In the present context, we limit ourselves to the type of resonances 
discussed in \citet{bu95,bu97} and \citet{go98} involving linear frequency 
combinations such that $\nu_1\,+\nu_2\,\sim\,2\,\nu_0$, and, more 
specifically, a particular case in which a $\ell=1$ mode is split by slow 
rotation and form a nearly symmetric triplet. This choice is obviously 
driven by the specific configuration of the modes studied in KIC~08626021, 
which, we recall, are all identified as $\ell=1$ rotationally split $g$-mode 
triplets. 

To clarify this further, we do not consider here other potential coupling 
mechanisms described, e.g., in \citet{wu01b} because they address a different 
problem, namely the problem of mode amplitude saturation through a proposed 
mechanism that indeed involves a nonlinear resonant mode coupling, but with 
one parent mode that is overstable (thus gaining energy) and two independent 
child modes that are damped (thus dissipating the energy). In our case, we 
observe and focus on a different nonlinear resonant coupling that 
occurs within triplets of l=1 $g$-modes resulting from the slow rotation 
of the star. The 3 modes in the triplets are overstable and nonlinearly 
interact with each other because slow rotation induces a near resonance 
relation between their frequencies (see below). 
\citet{wu01b}, \citet{wu01a}, \citet{mo05}, and other related studies do 
not treat this case and therefore cannot be helpful to describe and interpret 
what is occuring inside a rotationally split triplet. The only available 
framework for this is the \citet{bu95,bu97} and \citet{go98} papers 
that explicitely developed a theory to describe this kind of interaction 
and that should not be confused we various other works on nonlinear 
interactions between modes. 
We point out that it does not mean that the \citet{wu01b} mechanism
cannot also occur in KIC~08626021, but considering the linear growth rates 
expected for the observed modes (see Section 3.3 and Table 2), eventual 
limit cycles leading to cyclic amplitude variations would have timescales 
much longer ($\sim 1000$ yrs) than what is seen. This could hardly be 
connected to the observed features and would most likely not be noticeable 
in the available data that only cover a 2 year time baseline.}

{Going back to the configuration of interest involving rotationally 
split triplets,}  
rotation treated to first order approximation would 
lead to a strictly symmetric triplet that exactly satisfies the above 
mentionned relationship. However, terms of higher order are never exactly zero and a small asymmetry, 
dominated by the second order term, always exists. This asymmetry is in fact essential for driving 
the various resonant coupling behaviors.
The second order effect of rotational splitting, $\delta\nu_t$, that matters
can be estimated following the equation given in \citet{go98}:
\begin{equation}
\delta\nu_t = 4 C_{k\ell} \frac{\Omega^2}{\nu_0} 
\end{equation}
where $C_{k\ell}$ is the first order Ledoux constant ($\sim 0.5$ for dipole $g$-modes) 
and $\Omega\,=\,1/P_{rot}$ is the rotation frequency of the star. 
$P_{rot}$ is estimated from the first order average separation, $\Delta\nu$, 
between the components of the triplets and its value is $\sim 1.75$ days for
KIC 08626021 (see Section 2).
An asymmetry can also be evaluated directly from the measured frequencies 
of each triplet component, simply from the relation 
\begin{equation}
\delta\nu_o = \nu_- + \nu_+ - 2 \nu_0 \qquad .
\end{equation}

According to the resonant AEs from \citet{bu95} in which they ignored the 
slight interactions between modes with different $\ell$ and $k$,  
for the components in the $\ell=1$ triplet with frequencies $\nu_-$, $\nu_0$ and $\nu_+$, the corresponding 
amplitudes $A_-$, $A_0$ and $A_+$ and phases $\phi_-$, $\phi_0$ and $\phi_+$ should obey the following relations
\begin{subequations}
\begin{align}
  \frac{dA_-}{dt} & =  \kappa_-A_- + R_-A_0^2A_+ \cos(\Phi - \delta_-)  \notag \\
                  &  -A_-(q_{11}A_-^2 + q_{12}A_0^2 +q_{13}A_+^2)  \label{seq1}
\end{align}
\begin{align}
  \frac{dA_0}{dt} & = \kappa_0A_0 + R_0A_0A_+A_- \cos(\Phi + \delta_0)  \notag \\
                  &   -A_0(q_{21}A_-^2 + q_{22}A_0^2 +q_{23}A_+^2)  \label{seq2}
\end{align}
\begin{align}                  
   \frac{dA_+}{dt}& = \kappa_+A_+ + R_+A_0^2A_- \cos(\Phi - \delta_+)  \notag\\
                  &  -A_+(q_{31}A_-^2 + q_{32}A_0^2 +q_{33}A_+^2)  \label{seq3}    
\end{align}
\begin{align}
  \frac{d\Phi}{dt}& = \delta\nu - 2R_0A_-A_+ \sin(\Phi + \delta_0)  \notag\\
                  &  +A_0(R_-\frac{A_+}{A_-}\sin(\Phi - \delta_-)+R_+\frac{A_-}{A_+}\sin(\Phi - \delta_+)) \label{seq4} 
\end{align}
\end{subequations}
where $R_{-,0,+}$ and $q_{[(1,2,3)(1,2,3)]}$ are the nonlinear coupling coefficients 
associated to each component. Their values depend on complex integrals of the 
eigenfunctions of the modes involved in the coupling. The quantities $\kappa_-$, $\kappa_0$, and
 $\kappa_+$ are the linear growth rates of the $m=-1, 0, +1$ components, respectively. 

The numerical solutions of the AEs associated with this resonance {
identify three distinct regimes} (see the example provided in 
\citealp{bu97}). In order of 
magnitude, the occurrence of these three regimes can be roughly quantified 
by a parameter, $D$, defined as 
\begin{equation}
D \equiv \frac{2\pi\delta\nu}{\kappa_0}.
\end{equation}
{But the ranges for this parameter defining the boundaries of the 
various regimes depends somewhat on the values of the nonlinear 
coefficients in the real star}. 

The {first predicted situation} is the nonlinear frequency lock regime 
in which the 
observed frequencies appear in exact resonance ($\delta\nu=0$) and the 
amplitudes are constant. In the case of the DB white dwarf star GD358, 
numerical solutions of the AEs indicated that the range of the $D$ parameter 
corresponding to this regime was between 0 and 20 \citep{go98}. 
{However, these values are probably not universal and depend on the 
specific properties of the mode being considered, in particular on the 
value of the linear growth rate, $\kappa_0$, of the central component of 
the considered triplet}.  

When the triplet components move away from the resonance center ($\delta\nu\ne0$), 
they enter the so-called intermediate regime where amplitude and frequency 
are no longer stable and modulations can appear in the pulsations. In this 
regime, periodic variations can be expected with a timescale of 
\begin{equation}
P_{mod} \sim \frac{1}{\delta\nu}\simeq\frac{2\pi}{\kappa_0}\frac{1}{D}\qquad,
\end{equation}
{i.e., roughly the timescale derived from the inverse of the frequency 
asymmetry of the triplet (dominated by the second order effect of 
stellar rotation), which is connected to the inverse of the growth rate of 
the pulsating mode by the $D$ parameter \citep{go98}}.

Far from the resonance condition, the modes recover the regime of steady 
pulsations with nonresonant frequencies. In the nonresonant regime, 
the nonlinear frequency shifts become very small and the frequencies are 
close to the linear ones. 

We finally point out that in addition to the above three regimes, there 
exits a narrow hysteresis (transitory) regime between the frequency lock and 
intermediate regimes where the frequencies can be locked while the amplitudes 
still follow a modulated behavior.

\subsection{Connection with the observed triplets}

In light of the theoretical framework summarized above, {we point out 
that some of the behaviors observed in the 2 triplets $f_1$ and $f_2$ and 
in the doublet $f_3$ (an incomplete triplet) can be quite clearly 
connected to nonlinear resonant couplings occurring in different regimes}. 
We discuss each case below, 
but since the linear growth rate of the modes is an important ingredient
to these resonance mechanism, we provide first some results of linear 
nonadiabatic pulsation calculations specifically tuned for a model representing 
best the DBV star KIC 08626021.

\subsubsection{Nonadiabatic properties of the observed modes}

Following our re-analysis of the data obtained for KIC 08626021 with {\sl Kepler}, 
the recognition that 8 independent periods have to be considered for 
a detailed asteroseismic study (and not only 7 as used in BK14)
coupled with our present need for a realistic seismic model representation 
of the star to carry out a nonadiabatic study of the mode properties led us 
to attempt a new asteroseismic analysis for this object. The details of this 
seismic study -- a subject by its own that deserves a specific attention -- 
are fully reported in \citet{gi15}. The seismic solution obtained by 
\citet{gi15} for KIC 08626021 constitutes a major improvement over any of 
the fits proposed so far for this star, considering that it reproduces the 
8 independent periodicities to the actual precision of the {\sl Kepler} 
observations. It is therefore an excellent reference for our purposes. 

We used this specific seismic model to estimate the theoretical linear 
growth rates of the fitted pulsations modes. 
These computations were done using two different nonadiabatic pulsation
codes, one still working in the frozen convection (FC) 
approximation \citep{br92,fo94,br97}
and the other implementing a more realistic time-dependent convection 
(TDC) treatement \citep{du01,gr05}. In DA and DB white dwarf pulsators, 
the superficial 
convection layer has an important contribution to the driving of modes 
(through the sometimes called convective driving mechanism). The 
positions of the theoretical instability strips, in particular the blue edges, 
are particularly sensitive to the adopted treatment (TDC vs FC) and to 
the efficiency of convection itself that controls the depths of the 
convection zone (the $\alpha$ parameter in the Mixing Length Theory; 
see \citealp{van12}). These can also affect the growth rate of 
each individual mode. Unfortunately, the oscillation periods have 
essentially no sensitivity to the $\alpha$ parameter, which is therefore not 
constrained by seismology. In this context, we explored various combinations 
of $\alpha$ values for the two different nonadiabatic treatments of the 
convection perturbation to estimate the typical range of values one would 
expect for the growth rate of the modes. 

The results of these nonadiabatic calculations are summarized in Table~2 for 
the triplet (and doublet) components $f_1$, $f_2$, $f_3$, and, to be 
complete, for the other fitted frequencies as well. All these modes 
can effectively be driven in this star and the value of the growth rate 
$\kappa_0$ mostly depends on the radial order of the mode, strongly 
increasing when $k$ increases. For the modes of interest, we find that 
$\kappa_0$ lies in the ranges $2\cdot 10^{-13} - 7\cdot 10^{-12}$, 
$2\cdot 10^{-12} - 4\cdot 10^{-11}$, and $1\cdot 10^{-11} - 2\cdot 10^{-10}$
for $f_2$, $f_1$, $f_3$, which, from the seismic solution of
\citet{gi15}, are successive dipole modes of radial order $k=3$, 4, and 5, 
respectively.

\begin{table*} \caption[]{Mode identification and linear non adiabatic 
growth rates, $\kappa_0$, from the best seismic model of KIC 08626021.}
\begin{center}
\small
\begin{tabular}{lccccccccccl}  
\hline\\[-0.8em]
Id.  & Frequency & $\ell$ & $k$ & $C_{k\ell}$ & $\kappa_0^\dagger$ & $\delta\nu_t$ & $P_m$(th) & $\delta\nu_o$ & $P_m$(obs) & $D$ & Comment\\
     & ($\mu$Hz) &  & & & & ($\mu$Hz) & (day) & ($\mu$Hz) & (day) &  & \\
\hline\\[-0.8em]
 $f_{2,0}$ & 5073.23411 & 1 & 3 & 0.426 & $2.10^{-13} - 7.10^{-12}$  & 0.0148 & 780 & 0.0064 & $\sim 700$ & $1.10^4-5.10^5$ & Hysteresis regime\tablefootmark{$\star$}\\
 $f_{1,0}$ & 4309.91490 & 1 & 4 & 0.456 & $2.10^{-12} - 4.10^{-11}$  & 0.0187 & 620 & 0.00034 & $\sim 620$ &$3.10^3-6.10^4$ & Intermediate regime\\
 $f_{3,0}$ & 3681.80287 & 1 & 5 & 0.469 & $1.10^{-11} - 2.10^{-10}$  & 0.0223 & 518 & ... & 263 & $7.10^2-1.10^4$ & Intermediate regime\\
 &\\
 $f_6$ & 3294.36928 & 1 & 6 & 0.467 & $2.10^{-11} - 4.10^{-10}$  & ... & ... & ... & ... & ... & \\
 $f_9$ & 6981.26129 & 2 & 4 & 0.121 & $7.10^{-12} - 7.10^{-11}$  & ... & ... & ... & ... & ... & \\
 $f_5$ & 4398.37230 & 2 & 8 & 0.152 & $4.10^{-10} - 5.10^{-9}$   & ... & ... & ... & ... & ... & \\
 $f_7$ & 3677.99373 & 2 & 10& 0.154 & $2.10^{-9} - 4.10^{-8}$    & ... & ... & ... & ... & ... & \\
 $f_4$ & 2658.77740 & 2 & 15& 0.161 & $1.10^{-7} - 6.10^{-7}$    & ... & ... & ... & ... & ... & \\
 \hline
\end{tabular}
\normalsize
\label{tnadiab}
\end{center}
\tablefoot{
\tablefoottext{$\dagger$}{A range of values is given for the growth rate, reflecting calculations
using various treatments and hypotheses on the efficiency of convection (see text);}
\tablefoottext{$\star$}{The frequencies are locked but the amplitudes are still modulated.} 
}
\end{table*}

\subsubsection{triplets in the intermediate regime}

The periodic amplitude and frequency modulations observed in the triplet 
at 4310\,$\mu$Hz ($f_1$) immediately suggest that this triplet is in the intermediate 
regime of the resonance (see again Fig.\,\ref{pf_mod}). Both the prograde and 
retrograde components show a modulation of frequency and amplitude with a 
period of $\sim$\,600\,d. The central ($m=0$) component of $f_1$ has a 
frequency and amplitude modulation that is perhaps slightly longer 
($\sim$\,680\,d; precision is low here as this is about the same timescale 
as the duration of the observing campaign), but remains of the same order.
For comparison purposes, we provide in Table~2 the modulation timescale 
$P_m(\rm th)\sim 1/\delta\nu_t$ (Eqn. 5) expected from the asymmetry, $\delta\nu_t$, 
caused by the second order correction to the rotational splitting. The 
latter is computed with the $C_{k\ell}$ value obtained from the reference 
model and the value $P_{\rm rot} = 1.75$ days for the rotation period of 
the star. With $P_m(\rm th) \simeq 620$ days, the value obtained is 
sufficiently consistent with the observed modulation period to support 
the idea that we have indeed uncovered the right explanation for the 
behavior of the components in this triplet. Interestingly, the asymmetry 
can also be derived directly from the measured frequencies. 
Using directly the values given in Table~\ref{t1}, the quantity $\delta\nu_o$ 
represents the asymmetry for the frequencies averaged over the 
observation time baseline. We find it to be very small, i.e., much smaller 
than $\delta\nu_t$, suggesting that even in this intermediate regime
the nonlinear interactions may already have forced the frequencies of the 
triplet components to a locked position (where $\delta\nu\rightarrow 0$), on 
average (since the frequencies are still varying with time, oscillating 
around their mean value).
 
According to the nonlinear resonant coupling theory, all the three components in 
a triplet should have the same modulations, both in amplitude and frequency. 
The slight difference between the side components and the central component 
in terms of the modulation period might be that the interaction of the modes 
in the DBV star is more complex than the idealized case described by the theory. 
It might also be a suggestion that the growth rates for each component of the 
triplet are not similar (as is assumed in this theoretical framework). 
The shape of the amplitude modulations of the retrograde component $f_{1,-}$ is 
not as smooth as the other two components. This might be caused by the additional 
coupling of the mode with $f_4$ at 2659\,$\mu$Hz (see Section 2.4.4 and 
Figure\,\ref{c_mod}). Such a coupling occurring outside the triplet is not 
considered by \citet{bu95} who neglects other interactions with independent 
modes {(i.e, the triplet is considered as an isolated system)}.

The second structure that can {also} be associated to the intermediate 
regime is the doublet $f_3$. We recall that the best interpretation for 
this doublet 
is that it belongs to a $\ell=1$ triplet with one of the side components 
(the low frequency one, $f_{3,-}$) missing, most likely because its amplitude 
is below the detection threshold. The two remaining components show clear 
periodic modulations of both frequencies and amplitudes.
All variations occur on a somewhat shorter timescale of 
$P_m({\rm obs}) \sim 263$ days
(meaning that they are fully resolved in our data set, contrary to the 
modulations of $f_1$; see Section 2.4.3 and Fig.~\ref{tt_mod}), except 
for the frequency of the $f_{3,-}$ component whose modulation period 
appear to be approximately twice that value ($P_m({\rm obs}) \sim 550$ days). 
For this $f_3$ mode, the second order 
rotational splitting correction $\delta\nu_t$ also suggests a shorter 
modulation timescale of $P_m(\rm th) \sim 518$ days, which is 
comparable but not strictly identical. It is not possible in this case 
to evaluate $\delta\nu_o$ because of the missing third component.

\subsubsection{A triplet in the transitory hysteresis regime}

The case of the $f_2$ triplet at 5073 $\mu$Hz (see Section 2.4.2 and 
Figure\,\ref{st_mod}) is slightly different in that the frequencies are 
clearly stabilized while the amplitudes are modulated. This suggests that
$f_2$ is in another configuration, in between the frequency lock regime 
(where both amplitudes and frequencies are locked and therefore non-variable)
and the intermediate regime. {This configuration could be linked to 
the narrow} transitory hysteresis regime briefly mentionned in section 
3.1. This finding shows that two neighbor triplets can belong to 
different 
resonant regimes (frequency lock, narrow transition, intermediate or 
nonresonant), as it was also noticed for the white dwarf star GD 358 
\citep{go98}.

\subsection{Linear growth rates and the D parameter}

Table\,\ref{tnadiab} also provides the estimated values for the parameter 
$D$ derived from Eqn.~(4) and from the values of $\delta\nu_t$ (Eqn.~1) 
and $\kappa_O$ (obtained from the seismic model of KIC~08626021; see 
Section 3.2.1). We find that $D$ lies in the ranges $3.10^3-6.10^4$, $1.10^4-5.10^5$ 
and $7.10^2-1.10^4$ for the triplets $f_1$, $f_2$ and $f_3$, respectively. 
These values are at least one order of magnitude larger than the range given 
in \citet{go98} for the intermediate regime ($D\sim 9-25$ for the white dwarf 
star GD\,358). This large difference is clearly caused by the linear growth 
rates ($\kappa_0$) adopted for the modes. Our values come from a detailed 
linear nonadiabatic calculation based on the seismic model. Since the 
3 triplets are fitted to low radial order consecutive modes ($k=3$, 4, and 5), 
their corresponding linear growth rates are generally small and differ 
substantially from one mode to the other ($\kappa_0$ increases rapidly with $k$).
In contrast, \citet{go98} roughly scaled the growth rate of the modes according to 
the relationship $A_k^2/A^2_{k^{'}} \sim \kappa_k/\kappa_k^{'}$, assuming that 
all the coupling coefficients $q^R$ are of the same order of magnitude, 
leading to estimated values of $\kappa_0 \sim 10^{-8}$. Values comparable to
\citet{go98} for the growth rate could be obtained only if the 3 triplets 
were assigned to higher radial orders ($k$ between 10 and 15 instead of 
3 to 5). This would require a huge shift compared to the current seismic 
solution which is clearly not permitted on the seismic modeling side. 

In the AEs formalism of \citet{bu97}, the solutions admit three distinct 
regimes and one narrow transitory regime. Those regimes are related to 
the distance from the resonance center (i.e., $\delta\nu = 0$). The parameter 
$D$ in this transitory regime should be slightly smaller than in the intermediate 
regime, as this transitory regime is closer to the resonance center.
This means that $D$ should be smaller for $f_2$, which is in this transitory 
regime, compared to $f_1$ and $f_3$ that are in the intermediate regime. 
The ranges given for the $D$ values in Table\,\ref{tnadiab} still permit 
this constraint to be roughly satisfied, but the overall larger $D$ values 
for $f_2$ could also lead to a contradiction here.

We think at this stage that further quantitative comparisons between theoretical 
considerations and the observed properties of the modulations would require 
to solve the amplitude equations specifically for this case. {This is 
however beyond the scope of this paper, as no specific modeling 
tools} for these nonlinear effects is available to us at present. {We 
emphasize that with a detailed numerical solution} of the nonlinear 
amplitude equations, the unknown coupling coefficients could, in principle, 
be determined from fitting the observed frequency and 
amplitude modulations. These coefficients, if known, would then allow us 
to derive the $D$ parameter which is strongly related to the different 
regimes of the nonlinear resonances. With the determination of this 
parameter, a measurement of the growth rate of the oscillation 
modes would then possibly follow, leading for the first time to an 
independent estimation of the linear nonadiabatic growth rates of the modes 
and a direct test of the nonadiabatic pulsation calculations.


\section{Summary and conclusion}

Frequency and amplitude modulations of oscillation modes have been found 
in several rotationally split triplets detected in the DB pulsator 
KIC~08626021, thanks to the high quality and long duration photometric data 
obtained with the {\sl Kepler} spacecraft. These modulations show signatures 
pointing toward nonlinear resonant coupling mechanisms occurring between the 
triplet components. This is the first time that such signatures are 
identified so clearly in white dwarf pulsating stars, although hints of 
such effects had already been found from ground based campaigns in the past
(e.g., \citealt{va11}).

Reanalysing in detail the nearly 2 years of {\sl Kepler} photometry obtained 
for this star, we have detected 13 very clear independent frequencies above 
our estimated "secure" detection threshold ($5.6\sigma$; see Section 
2.2 and Table~\ref{t1}), two frequencies that appear to be linear combinations 
of other independent modes, and two additional, but significantly less 
secured, frequencies emerging just above 5$\sigma$ the mean noise level. 
Overall, we find that our secured frequencies are consistant with those 
reported in BK14, but we somewhat differ on the interpretation of 
some structures in the frequency spectrum.

Most notably, we find that 3 frequencies in the 3677$-$3686 $\mu$Hz range, 
formerly identified as the components of a single triplet by BK14, 
cannot be interpreted like this. We conclude instead that one of the 
frequency ($f_7$ in Table~\ref{t1}) is an independent mode while the 
two others ($f_{3,0}$ and $f_{3,+}$) form the two visible components of 
an incomplete triplet whose third component is not seen. This has some 
implications for the seismic modeling which should in fact include 8 
independent frequencies and not only 7 as in BK14. A new 
detailed seismic analysis of KIC~08626021 based on these 8 modes is 
provided by \citet{gi15}. The frequency spacings (observed between the two 
components of $f_3$ and the components of two other well identified triplets,
$f_1$ and $f_2$) indicate an average rotation period of $\sim 1.75$ days 
for KIC~08626021, i.e., in agreement with the value given by BK14.

Also differing from BK14, we find that the 2 components of the 
$f_3$ doublet have a "hyperfine" structure with sidelobes separated by
0.044 $\mu$Hz, indicating a modulating phenomenon occurring on a long 
time scale of $\sim 263$ days. {In addition, the components 
forming triplets show long term (quasi)-periodic frequency and/or 
amplitude modulations} that appear to be correlated, as they evolve either 
in phase or antiphase.
{The triplet at 4310\,$\mu$Hz ($f_1$) show signs of periodic 
modulation} of both the frequencies and amplitudes with a timescale of 
roughly 600 days with the side components evolving in phase, while the 
central mode is in antiphase. {The timescale appears somewhat} shorter
(263 days) for the $f_3$ doublet while the triplet $f_2$ shows only 
modulations in amplitudes (the frequencies appear stable during the 
observations) {with a probable timescale of $\sim 700$ days}.

We show that these behaviors can be related to the so-called 
nonlinear resonant coupling mechanisms that is expected to occur 
within rotationnally split triplets. The amplitude equations 
\citep{bu97,go98} predict 3 main regimes in which the triplet components 
may behave differently. {It appears that $f_1$ and $f_3$ can be 
linked to the so-called intermediate regime} of the resonance where 
both the amplitude and frequency of the modes {should} experience a 
periodic modulation. 
We find that {the timescales expected from the theory are quite
consistent with the observed periodicities of the modulations}.
The triplet $f_2$ shows a different behavior that can be associated with
a narrow transitory hysteresis regime between the intermediate regime and 
the frequency locked regime in which locked frequencies and modulated 
amplitude solutions can coexist.

We also found correlated frequency and amplitude modulations in a linear 
combination of frequencies involving the modes $f_{1,-}$ and $f_4$, and 
the frequency $f_8$. This configuration may be related to parents/child mode 
interactions in pulsating stars, $f_8$ being either a combination 
frequency resulting from strong nonlinearities or, because the 
amplitude ratio is large, an eigenmode 
whose amplitude has been significantly enhanced by a resonant coupling 
phenomenon (see \citealt{br14}). Further investigations need to be 
carried out to evaluate which explanation is the most plausible.

As an additional step toward comparing more quantitatively observations to 
the theoretical expectations, we estimated theoretical linear growth rates 
(see Table~\ref{tnadiab}) of the triplet central components using the seismic 
model provided by \citet{gi15}. We used two different nonadiabatic pulsation 
codes for these computations: one working in the frozen convection approximation 
\citep{br92,fo94,br97} and the other implementing a time-dependent 
convection treatement \citep{du01,gr05}. 
The modes of interest $f_2$, $f_1$ and $f_3$ have growth rates that 
are in the ranges $2\cdot 10^{-13} - 7\cdot 10^{-12}$, 
$2\cdot 10^{-12} - 4\cdot 10^{-11}$, and $1\cdot 10^{-11} - 2\cdot 10^{-10}$,
respectively. With these values, we finally estimate the parameter $D$ 
(a key parameter that measures how far away is the mode from the 
resonance center) which is found in the range $3.10^3-6.10^4$, 
$1.10^4-5.10^5$ and $7.10^2-1.10^4$ for the mode triplet $f_1$, $f_2$ and $f_3$, 
respectively. These values are significantly larger than those estimated in 
\citet{go98} and need further investigation, but going beyond this 
would require to solve the amplitude equations for the specific case of 
KIC\,08626021, which is currently not possible. 

We want also to emphasize the fact that the uncovered frequency 
modulations, which are related to nonlinear coupling mechanisms and that 
occur on times scales long enough to be difficult to detect but short 
compared to the secular evolution timescale, can potentially impair any attempt
to measure reliably the effects of the cooling of the white dwarf on the 
pulsation periods. Measuring the changing rate of the pulsation periods in 
white dwarf stars could indeed offer an opportunity to constrain the neutrino emission 
physics \citep{wi04,su08}. However, one should be extremely careful 
of the potential contamination of nonlinear effects, which may need to be 
corrected first. Some independent modes in KIC\,08626021 that seem to be stable 
in frequency over much longer timescales and that do not apparently couple 
with other modes in the white dwarf star could be good candidates for measuring 
period rates of change. But nonlinear interactions could still be affecting 
them on longer timescales that we cannot detect with {\sl Kepler}.    

Finally, the observed periodic frequency and amplitude modulations that 
occur in the intermediate regime of the resonance may allow for new 
asteroseismic diagnostics, providing in particular a way to measure for the 
first time linear growth rates of pulsation modes in white dwarf stars. 
This prospect should motivate further theoretical work on nonlinear resonant 
mode coupling physics and revive interest in nonlinear stellar pulsation 
theory in general.

\begin{acknowledgements} 
Funding for the {\sl Kepler} mission is provided 
by NASA's Science Mission Directorate. We greatfully acknowledge the 
{\sl Kepler} Science Team and all those who have contributed to making 
the {\sl Kepler} mission possible. WKZ acknowledges the financial support 
from the China Scholarship Council. V. Van Grootel is an F.R.S-FNRS 
Research Associate. This work was supported in part by the 
Programme National de Physique Stellaire (PNPS, CNRS/INSU, France) and 
the Centre National d'Etudes Spatiales (CNES, France).
\end{acknowledgements}

\end{document}